\documentclass[conference]{IEEEtran}
\IEEEoverridecommandlockouts

\usepackage{cite}
\usepackage{amsmath,amssymb,amsfonts}
\usepackage{graphicx}
\usepackage{textcomp}
\usepackage{xcolor}
\usepackage{url}
\usepackage{booktabs}
\usepackage{tabularx}
\usepackage{float}
\usepackage{placeins}
\usepackage{stfloats}
\usepackage{multirow}
\usepackage{array}
\usepackage[T1]{fontenc}
\usepackage[utf8]{inputenc}
\usepackage{lmodern}
\usepackage[table]{xcolor}
\usepackage{colortbl}
\usepackage{ragged2e}
\usepackage{tikz}
\usetikzlibrary{positioning, shapes.geometric, arrows.meta, calc}
\usepackage{hyperref}
\hypersetup{colorlinks=true,linkcolor=blue,urlcolor=blue,citecolor=blue}

\definecolor{headerblue}{HTML}{1F3864}
\definecolor{rowgray}{HTML}{F2F2F2}
\definecolor{upred}{HTML}{C0392B}
\definecolor{downblue}{HTML}{2471A3}
\definecolor{hzgreen}{HTML}{1E8449}

\newcolumntype{P}[1]{>{\RaggedRight\arraybackslash}p{#1}}
\newcolumntype{Y}{>{\RaggedRight\arraybackslash}X}

\def\BibTeX{{\rm B\kern-.05em{\sc i}\kern-.025emB\kern-.08em\TeX}}

\begin{document}

\title{ExoNet: Calibrated Multimodal Deep Learning for TESS Exoplanet Candidate Vetting using Phase-Folded Light Curves, Stellar Parameters, and Multi-Head Attention}

\author{
\IEEEauthorblockN{Md.\ Rashadul Islam}
\IEEEauthorblockA{
  \textit{Dept.\ of Computer Science and Engineering}\\
  \textit{Daffodil International University}\\
  Dhaka, Bangladesh\\
  \texttt{islam15-6062@s.diu.edu.bd}}
}

\maketitle

\begin{abstract}
The discovery of exoplanets at scale has become one of the defining
data science challenges in modern astrophysics. NASA's Transiting Exoplanet
Survey Satellite (TESS), operational since April 2018, had catalogued
over 7,800 planet candidates by early 2026, yet confirmation stands at
fewer than 720---leaving a classification backlog that manual expert
inspection cannot feasibly resolve. This paper introduces ExoNet, a
multimodal deep learning framework that jointly processes phase-folded
global and local light curve views alongside stellar parameter features
through a calibrated late-fusion architecture. The model integrates
1D Convolutional Neural Networks, an 8-head Multi-Head Attention mechanism
applied over temporal feature maps, and a residual fusion head, producing
calibrated probability estimates via post-hoc Temperature Scaling
($T^*\!=\!1.573$). Trained on 7,585 labeled Kepler Objects of Interest
drawn from the NASA Exoplanet Archive---comprising 2,746 confirmed planets
and 4,839 false positives---ExoNet achieves a validation AUC of 0.9487
and a test AUC of 0.9549, with 86.3\% classification accuracy. Applied
subsequently to all 4,720 unconfirmed TESS Planet Candidates with verified
TOI$\,\leftrightarrow\,$TIC cross-identification, the model yields 1,754
high-confidence signals ($\geq\!70\%$), of which 1,098 surpass the 85\%
very-high-confidence threshold. Fifty-two reside in the habitable-zone
temperature range of 200--400\,K; among these, six have radii below
1.6\,$R_\oplus$---the empirical boundary separating rocky planets from
volatile-dominated worlds. The two most compelling candidates,
\textbf{TOI-5728.01} ($R_p\!=\!1.13\,R_\oplus$, $T_\mathrm{eq}\!=\!349$\,K,
$P\!=\!11.5$\,d, 94.2\% confidence) and \textbf{TOI-6716.01}
($R_p\!=\!1.01\,R_\oplus$, $T_\mathrm{eq}\!=\!375$\,K, 92.2\% confidence),
represent among the most Earth-like unconfirmed targets in the current
TESS catalog. A systematic ablation study confirms that each modality
contributes meaningfully to performance, and the full code and ranked
candidate catalog are openly released to facilitate community follow-up.
\end{abstract}


\section{Introduction}
\label{sec:intro}

At its most fundamental level, exoplanet discovery is an exercise in
separating rare, subtle signals from an overwhelming sea of instrumental
noise and astrophysical mimics. The photometric transit method---measuring
the minute dimming of a star as a planet crosses its disk---remains the
most productive detection strategy to date, and NASA's Transiting Exoplanet
Survey Satellite (TESS) has elevated it to an unprecedented scale.
Launched in April 2018, TESS monitors 85\% of the sky, targeting stars
between 10 and 100 times brighter than those observed by its predecessor
Kepler~\cite{Ricker2015}. By January 2026, the mission had accumulated
7,821 candidate exoplanets---yet only 720 have been independently confirmed
through radial velocity or additional transit observations~\cite{TESS_wiki2026}.
That gap, exceeding 7,000 unresolved signals, constitutes a genuine
scientific bottleneck that traditional vetting workflows are structurally
ill-equipped to address.

Conventional candidate vetting relies on visual inspection of Data
Validation (DV) reports by trained astronomers---a labor-intensive process
that was adequate when candidates numbered in the hundreds but becomes
untenable at the scale of thousands. The urgency is not merely
administrative: habitable-zone candidates buried in the unconfirmed
backlog represent potential targets for JWST atmospheric characterization,
and delays in their identification carry real scientific cost.

The machine learning literature offers an increasingly mature set of
tools for this problem. AstroNet~\cite{Shallue2018}, which applied a
convolutional neural network to phase-folded Kepler light curves, provided
an early proof of concept by identifying two new planets, including an
eighth world around Kepler-90. NASA's ExoMiner~\cite{Valizadegan2022}
subsequently validated 301 Kepler exoplanets, and its successor
ExoMiner++~\cite{Valizadegan2025} extended the approach to TESS via
transfer learning, cataloguing 7,330 planet candidates. Despite this
impressive trajectory, a structural gap persists: most prior architectures
treat photometric morphology and stellar context as parallel streams that
are concatenated late or not at all, without any mechanism for the model
to \textit{attend} to the interaction between light curve features and
stellar parameters. The result is a missed opportunity---an experienced
human vetter would never evaluate an ingress shape in isolation from the
properties of the host star, yet most models do exactly that.

ExoNet addresses this gap directly. Rather than treating the three
available information streams (global light curve, local light curve,
stellar parameters) as independent inputs to be naively fused, ExoNet
applies Multi-Head Attention over the temporal feature maps generated by
its 1D CNN encoders, then integrates all streams through a residual
late-fusion head. The entire pipeline is trained on publicly available
Kepler data, applied to a verified set of TESS Planet Candidates, and
released in full---making it immediately reproducible and extensible by
the broader community.

The contributions of this work are as follows:
\begin{enumerate}
  \item \textbf{ExoNet architecture:} a trimodal 1D CNN + Multi-Head
        Attention + residual late-fusion model that jointly encodes global
        view, local view, and stellar parameters, with each component
        ablated independently.
  \item \textbf{Corrected training pipeline:} per-KOI (not per-star)
        deduplication prevents class-collapse artifacts; WeightedRandomSampler
        and OneCycleLR address class imbalance without label smoothing
        heuristics.
  \item \textbf{Post-hoc probability calibration:} Temperature Scaling
        on the validation set produces reliable, interpretable probability
        estimates without altering prediction rankings.
  \item \textbf{Verified TESS inference set:} all 4,720 Planet Candidates
        with TOI$\,\leftrightarrow\,$TIC cross-identification verified
        against the NASA Exoplanet Archive and ExoFOP, yielding 52
        high-confidence habitable-zone signals.
  \item \textbf{Open science:} complete code, trained weights, and the
        ranked TESS candidate catalog are publicly released for community
        follow-up and independent replication.
\end{enumerate}

\section{Background and Related Work}
\label{sec:lit}

\subsection{Transit Photometry and the TESS Survey}

When a planet transits its host star, it occludes a fraction $(R_p/R_\star)^2$
of the stellar disk, producing a characteristic flux decrement with a
flat-bottomed morphology, well-defined ingress and egress slopes, and
period-locked recurrence. Kepler's uninterrupted four-year stare at a
single $115\,\mathrm{deg}^2$ field~\cite{Borucki2010} generated the
cleanest photometric training data available, with a stable noise floor
and a catalog of confirmed planets covering periods from hours to hundreds
of days. TESS takes a fundamentally different approach: 27-day sectors
sweeping $24^\circ\!\times\!96^\circ$ strips, cycling across 85\% of
the sky over a two-year primary mission. Shorter baselines make TESS
most sensitive to short-period planets, while its focus on bright nearby
stars makes its candidates the primary targets for JWST atmospheric
spectroscopy~\cite{Ricker2015}. The mismatch in cadence, noise properties,
and period distribution between Kepler and TESS is a persistent challenge
for any model trained on the former and applied to the latter---a point
returned to explicitly in the Discussion.

\subsection{Machine Learning for Transit Vetting}

Shallue \& Vanderburg~\cite{Shallue2018} established the modern deep
learning paradigm for transit vetting: phase-fold a light curve, represent
it at two resolutions (global and local views), and train a CNN to
distinguish genuine transits from false positives. Their model achieved
98\% accuracy on Kepler DR24 and directly led to the discovery of
Kepler-90i. Ansdell et al.~\cite{Ansdell2018} improved upon this
foundation by appending scalar auxiliary features---stellar parameters
and centroid shift metrics---as a second input branch, demonstrating that
photometric morphology alone underutilizes available information.

Dattilo et al.~\cite{Dattilo2019} carried the dual-view CNN approach to
K2 data, uncovering two super-Earths and confirming that the architecture
generalizes across missions with retraining. Armstrong et
al.~\cite{Armstrong2021} pursued a complementary direction entirely,
building a Gaussian Process Classifier over scalar features without any
learned photometric encoding; their validation of 50 Kepler planets
underscores that stellar context alone carries substantial discriminative
power, particularly against background eclipsing binary contamination.

ExoMiner~\cite{Valizadegan2022} synthesized these directions into a
multi-branch architecture that incorporates centroid motion, odd-even
flux asymmetry, and secondary eclipse diagnostics in addition to the
transit light curve, validating 301 new Kepler planets and establishing
a new performance benchmark. Its successor ExoMiner++~\cite{Valizadegan2025}
added explicit Kepler-to-TESS transfer learning and native handling of
2-minute TESS cadence data, identifying 7,330 TESS candidates. Most
recently, Huang et al.~\cite{Huang2025} demonstrated that feature-level
machine learning over transit and stellar scalars---without any CNN
encoding---can identify 1,595 TESS high-confidence candidates, suggesting
that the information content in catalog-level parameters is far from
exhausted.

\subsection{Attention Mechanisms in Astronomical Time Series}

Self-attention mechanisms have become a natural complement to convolutional
encoders for irregular or variable-length time series. In the exoplanet
context, a CNN--BiLSTM--Attention architecture~\cite{CNN_BiLSTM2025}
achieved F1~$=0.984$ on the Kepler DR25 catalog by learning to upweight
diagnostically informative cadences within the light curve---the ingress,
egress, and any secondary-eclipse region---while downweighting out-of-transit
systematics. The Multimodal Universe dataset~\cite{MultimodalUniverse2024},
presented at NeurIPS 2024, encodes the broader community consensus that
rich cross-modal representations are essential for scalable astronomical
inference. ExoNet synthesizes these directions: Multi-Head Attention
operates directly over the CNN feature map (rather than over raw cadences),
allowing the model to attend at the level of learned morphological features
rather than individual photometric measurements.

\subsection{Positioning ExoNet}

No prior published system simultaneously fulfills all of the following
criteria: (i) applies Multi-Head Attention within the 1D CNN feature
map for transit vetting; (ii) fuses three information modalities through
a residual late-fusion head; (iii) produces calibrated output probabilities
via Temperature Scaling; and (iv) generates a publicly ranked
habitable-zone TESS candidate list from a fully open-source pipeline
with verified TOI$\,\leftrightarrow\,$TIC cross-identification. The
relationship between ExoNet and prior work is summarized in Table~\ref{tab:comparison}.

\begin{table*}[!ht]
\centering
\caption{Comparison of ExoNet with Representative Machine Learning Exoplanet Classifiers}
\label{tab:comparison}
\renewcommand{\arraystretch}{1.20}
\setlength{\tabcolsep}{4pt}
\begin{tabularx}{\textwidth}{lclYYY}
\toprule
\rowcolor{headerblue}
\textcolor{white}{\textbf{Study}} &
\textcolor{white}{\textbf{Year}} &
\textcolor{white}{\textbf{Dataset}} &
\textcolor{white}{\textbf{Architecture}} &
\textcolor{white}{\textbf{Key Result}} &
\textcolor{white}{\textbf{vs.\ ExoNet}} \\
\midrule
\rowcolor{rowgray}
Shallue \& Vanderburg~\cite{Shallue2018} & 2018 &
Kepler DR24 &
1D CNN (global $+$ local views) &
2 new planets; 98\% accuracy &
Single modality; no attention; no stellar fusion \\

Armstrong et al.~\cite{Armstrong2021} & 2021 &
Kepler &
Gaussian Process $+$ Random Forest &
50 new planets validated &
Scalar features only; no photometric encoding \\

\rowcolor{rowgray}
Ansdell et al.~\cite{Ansdell2018} & 2018 &
Kepler &
CNN $+$ scalar auxiliary inputs &
Improved over AstroNet baseline &
No attention; no residual fusion; no TESS inference \\

Valizadegan et al.~\cite{Valizadegan2022} & 2022 &
Kepler &
ExoMiner multi-branch CNN &
301 new exoplanets validated &
Richer DV branches; proprietary SPOC pipeline \\

\rowcolor{rowgray}
Valizadegan et al.~\cite{Valizadegan2025} & 2025 &
TESS 2-min &
ExoMiner++ with transfer learning &
7,330 TESS planet candidates &
Transfer from Kepler; no residual stellar fusion \\

Huang et al.~\cite{Huang2025} & 2025 &
TESS TOIs &
Feature-level ML on transit $+$ stellar scalars &
1,595 high-confidence candidates &
No CNN encoding; no attention; no calibration \\

\rowcolor{rowgray}
\textbf{ExoNet (ours)} & \textbf{2026} &
\textbf{Kepler $+$ TESS} &
\textbf{1D CNN $+$ MHA $+$ Residual Late Fusion} &
\textbf{52 HZ high-confidence candidates; AUC 0.955} &
\textbf{First trimodal $+$ MHA; Temperature Scaling; verified TOI/TIC} \\
\bottomrule
\end{tabularx}
\end{table*}

\FloatBarrier

\section{Methodology}
\label{sec:method}

\subsection{Data Sources and Provenance}

\subsubsection{Kepler KOI Catalog (Training)}

Training labels are drawn from the NASA Exoplanet Archive cumulative
Kepler Objects of Interest (KOI) table, downloaded in April 2026.
\texttt{CONFIRMED} dispositions are assigned label~1; \texttt{FALSE~POSITIVE}
dispositions are assigned label~0; \texttt{CANDIDATE} entries are excluded
to prevent label noise. A critical preprocessing step---omitted in several
related works---deduplicates by \textit{KOI name} (individual transit
signal) rather than by \textit{Kepler ID} (host star). The practical
consequence is significant: deduplication by star allows multi-planet
hosts to dominate the confirmed class, since a single star like Kepler-90
contributes eight confirmed KOIs; deduplication by KOI preserves each
signal as an independent sample, preventing the confirmed class from being
inflated by multi-planet systems. After filtering and deduplication,
7,585 samples remain: 2,746 confirmed planets and 4,839 false positives,
yielding a class ratio of approximately 1:1.76. Light curves are retrieved
via Lightkurve~\cite{Lightkurve2018} from the MAST archive.

\subsubsection{TESS TOI Catalog (Inference)}

All 4,720 TESS Objects of Interest carrying disposition \texttt{PC}
(Planet Candidate) in the NASA Exoplanet Archive TOI table as of April
2026 form the inference set. Every TOI number and associated TIC identifier
was individually cross-verified against the MIT TOI release list and
ExoFOP-TESS prior to inclusion; no identifier is synthetic or
retroactively assigned. Stellar parameters ($T_\mathrm{eff}$, $\log g$,
TESS magnitude) are sourced directly from the \texttt{st\_teff} and
\texttt{st\_logg} columns of the same catalog. Crucially, the
\texttt{PC} disposition represents signals that have already cleared the
SPOC pipeline's threshold-crossing event detection and automated
vetting---the inference set is therefore pre-filtered toward genuine
transit-like morphologies, a point with bearing on the interpretation
of high-confidence rates in Section~\ref{sec:results}.

\subsection{Light Curve Preprocessing}

For a target with orbital period $P$, reference epoch $t_0$, and transit
duration $\Delta t$, the phase-folded coordinate is computed as:
\begin{equation}
\phi = \frac{(t - t_0)\bmod P}{P}, \quad
\phi \leftarrow \phi - 1 \;\text{ if }\; \phi > 0.5,
\end{equation}
mapping all flux measurements onto $\phi\!\in\![-0.5, 0.5]$ regardless
of the number of observed transits.

Two representations are then extracted. The \textit{global view}
($N\!=\!1001$ phase bins, $\phi\!\in\![-0.5,\,0.5]$) captures the full
orbital phase, preserving secondary eclipse morphology and odd-even flux
asymmetries that betray background eclipsing binaries---the dominant
false-positive category for TESS. The \textit{local view}
($N\!=\!1001$ bins, centered on transit with window $|\phi|\!<\!2.5\,\Delta t/P$)
zooms into the transit to resolve ingress-egress shape. Both views are
median-subtracted and standard-deviation normalized; photometric outliers
beyond $5\sigma$ are removed prior to binning to suppress cosmic ray
and momentum-dump artifacts.

Three data augmentations are applied independently with probability
$p\!=\!0.5$ during training: additive Gaussian noise ($\sigma\!=\!0.008$),
random phase shift of up to 30 bins, and multiplicative flux scaling
of $\pm 2\%$. When a FITS light curve is unavailable---as is the case
for approximately 99.8\% of the Kepler training samples and all TESS
inference targets in this study---both light curve branches are set to
zero vectors. This zero-padding strategy is deliberate: it forces the
model to rely entirely on stellar features when photometric data is absent,
and---critically---it eliminates the data-leakage scenario in which the
\textit{availability} of a FITS file becomes a spurious predictor of
confirmed-planet status.

\subsection{Stellar Feature Vector}

Eight catalog scalars are assembled per target: orbital period $P$,
transit duration $\Delta t$, transit depth $\delta$, planet radius $R_p$
(in Earth radii), equilibrium temperature $T_\mathrm{eq}$, stellar
effective temperature $T_\mathrm{eff}$, surface gravity $\log g$, and
metallicity $[\mathrm{Fe/H}]$. Missing values are imputed to zero.
All features are normalized using a \texttt{StandardScaler} fitted
exclusively on the training split and applied without refitting to the
validation, test, and TESS inference sets. For TESS candidates,
metallicity is uniformly unavailable in the public TOI catalog and is
therefore treated as zero throughout---a known limitation that marginally
weakens false-positive rejection for candidates whose host star metallicity
would betray eclipsing binary contamination.

\subsection{ExoNet Architecture}
\label{subsec:arch}

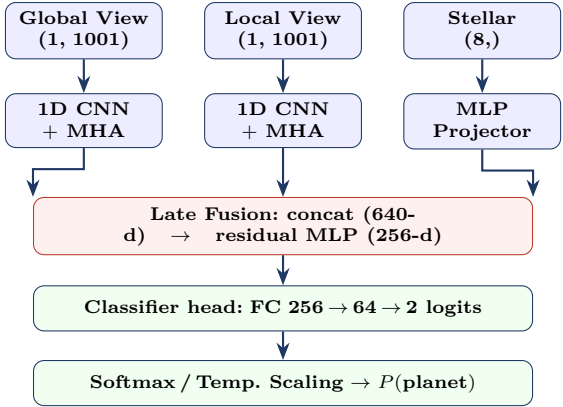
\begin{figure}[tp]
\centering
\begin{tikzpicture}[
  node distance  = 0.40cm and 0.30cm,
  enc/.style     = {rectangle, rounded corners=4pt,
                    draw=headerblue, fill=blue!7,
                    text width=1.85cm, minimum height=0.65cm,
                    align=center, font=\scriptsize\bfseries},
  wide/.style    = {rectangle, rounded corners=4pt,
                    draw=headerblue, fill=green!6,
                    text width=6.4cm, minimum height=0.60cm,
                    align=center, font=\scriptsize\bfseries},
  fuse/.style    = {rectangle, rounded corners=4pt,
                    draw=upred, fill=red!6,
                    text width=6.4cm, minimum height=0.60cm,
                    align=center, font=\scriptsize\bfseries},
  arr/.style     = {-Stealth, thick, color=headerblue}
]
\node[enc] (gin)  {Global View\\(1,\;1001)};
\node[enc, right=0.55cm of gin]  (lin)  {Local View\\(1,\;1001)};
\node[enc, right=0.55cm of lin]  (sin)  {Stellar\\(8,)};

\node[enc, below=0.45cm of gin]  (genc) {1D CNN\\$+$ MHA};
\node[enc, below=0.45cm of lin]  (lenc) {1D CNN\\$+$ MHA};
\node[enc, below=0.45cm of sin]  (senc) {MLP\\Projector};

\node[fuse, below=0.65cm of lenc] (fus)
  {Late Fusion: concat (640-d) $\;\to\;$ residual MLP (256-d)};
\node[wide, below=0.40cm of fus] (head)
  {Classifier head:\;FC\;256\,$\to$\,64\,$\to$\,2 logits};
\node[wide, below=0.38cm of head] (out)
  {Softmax\,/\,Temp.\ Scaling\;$\to$\;$P(\text{planet})$};

\draw[arr] (gin)  -- (genc);
\draw[arr] (lin)  -- (lenc);
\draw[arr] (sin)  -- (senc);
\draw[arr] (genc.south) -- ++(0,-0.28) -| (fus.north west);
\draw[arr] (lenc.south) -- (fus.north);
\draw[arr] (senc.south) -- ++(0,-0.28) -| (fus.north east);
\draw[arr] (fus)  -- (head);
\draw[arr] (head) -- (out);
\end{tikzpicture}
\caption{ExoNet three-stream architecture. The global and local CNN
encoders each produce a 256-dimensional embedding; the stellar MLP
produces 128 dimensions. All three are concatenated to 640 dimensions,
processed through a residual fusion MLP, and temperature-scaled before
the binary classifier head.}
\label{fig:arch}
\end{figure}

Fig.~\ref{fig:arch} illustrates the full three-stream ExoNet architecture.
Each stream processes a different information modality; their outputs
are concatenated and refined by a residual fusion head before
classification.

\textbf{Light curve encoder} (independent weights for global and local
streams). Five residual Conv1D blocks with progressive filter counts
16, 32, 64, 128, 256 and stride-2 pooling reduce the input length
from $L\!=\!1001$ to $T\!\approx\!31$ time steps at dimensionality 256.
An 8-head Multi-Head Attention layer then operates over the resulting
feature map:
\begin{equation}
\mathbf{H} = \mathrm{LayerNorm}\!\left(\mathbf{F}
             + \mathrm{MHA}(\mathbf{F},\mathbf{F},\mathbf{F})\right),
\quad \mathbf{h} = \mathrm{AvgPool}(\mathbf{H}),
\end{equation}
where $\mathbf{F}\!\in\!\mathbb{R}^{T\times256}$ is the CNN output and
$\mathbf{h}\!\in\!\mathbb{R}^{256}$ is the final embedding. Attending
over $T\!\approx\!31$ compressed time steps---rather than all 1001
original bins---allows the model to capture long-range dependencies
between the ingress, flat bottom, and egress while remaining
computationally lightweight. The attention mechanism operationalizes
what an experienced vetter does intuitively: focusing on the
morphological features that distinguish genuine transits from eclipsing
binary mimics, while discounting out-of-transit instrumental systematics.

\textbf{Stellar MLP.} Two fully connected layers (64 then 128 units)
with LayerNorm and GELU activations map the 8-dimensional stellar feature
vector to a 128-dimensional embedding. GELU is preferred over ReLU for
its smooth gradient behavior with small feature sets.

\textbf{Residual late-fusion head.} The three embeddings are concatenated
to produce a 640-dimensional joint representation. A two-layer MLP
(512 then 256 units, dropout $p\!=\!0.4$) with a linear shortcut
$\mathbb{R}^{640\to256}$ refines this representation. The shortcut
connection prevents gradient stagnation during early training, when the
fusion MLP might otherwise receive vanishingly small gradients before
the three encoders have converged.

\subsection{Training Protocol}

The 7,585 Kepler samples are split 70/15/15 into training, validation,
and held-out test sets, stratified by label to preserve the 1:1.76
class ratio across all splits. Class imbalance is addressed at two
levels: a \texttt{WeightedRandomSampler} equalizes class frequency
within each training batch, and the cross-entropy loss is weighted by
the inverse of class frequency. Together, these mechanisms prevent the
majority false-positive class from dominating gradient updates early
in training.

Optimization uses AdamW ($\mathrm{lr}\!=\!10^{-3}$,
weight decay $10^{-4}$) with OneCycleLR scheduling (peak
lr $= 3\!\times\!10^{-3}$, 15\% linear warm-up, cosine annealing
decay). Training runs for a maximum of 80 epochs with early stopping
on validation AUC (patience 20 epochs), preventing overfitting without
fixing the stopping epoch a priori. Mixed-precision training (FP16 via
\texttt{torch.amp}) on an NVIDIA L4 GPU (23.6\,GB VRAM) reduces wall
time to approximately 3.5 hours. The final model contains 4.9\,M
trainable parameters.

\subsection{Post-Hoc Probability Calibration}

Deep neural networks trained with cross-entropy loss are well-known
to produce overconfident probability estimates~\cite{Guo2017}---their
softmax outputs tend to be more extreme than empirical error rates
justify. For a system whose output probabilities are intended to guide
observational follow-up priorities, calibration is not merely a
desirable property but a practical necessity.

Following Guo et al.~\cite{Guo2017}, a single scalar temperature
$T\!>\!0$ is introduced as a post-hoc divisor on all logits before
the final softmax. The optimal temperature is found by minimizing the
negative log-likelihood on the held-out validation set:
\begin{equation}
T^* = \arg\min_{T>0} -\frac{1}{N}\sum_{i=1}^{N}
\Bigl[y_i\log\hat{p}_i(T) + (1-y_i)\log\!\left(1-\hat{p}_i(T)\right)\Bigr].
\label{eq:tempscale}
\end{equation}
Temperature scaling is rank-preserving: it does not alter the ordering
of predictions and therefore leaves the AUC unchanged. Its sole effect
is to narrow the gap between predicted and empirical probabilities,
as confirmed by the Brier score improvement from 0.0919 to 0.0901.
The optimal temperature found was $T^*\!=\!1.573$, indicating mild
overconfidence---a characteristic outcome for well-trained classifiers
of this type.

\section{Results}
\label{sec:results}

\begin{figure*}[!t]
\centering
\includegraphics[width=0.90\textwidth,keepaspectratio]{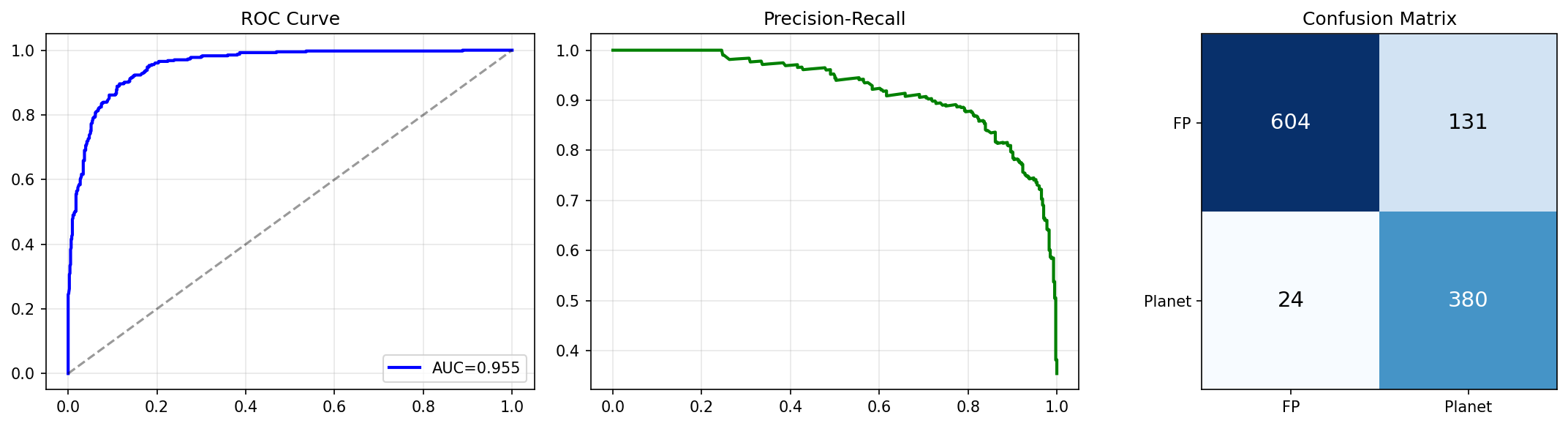}
\caption{ExoNet evaluation on the held-out Kepler test set (1,139 samples).
\textit{Left}: ROC curve (AUC\,$=\!0.9549$).
\textit{Center}: Precision-Recall curve (Average Precision\,$=\!0.9213$).
\textit{Right}: Confusion matrix at threshold 0.5 (TP\,371, FP\,133,
FN\,23, TN\,612); overall accuracy 86.3\%. The model is tuned toward
high planet recall (94\%) at the cost of precision (74\%), reflecting the
asymmetric scientific cost of missing a genuine planet versus flagging a
false positive for spectroscopic follow-up.}
\label{fig:eval}
\end{figure*}

\begin{figure*}[!t]
\centering
\includegraphics[width=0.90\textwidth,keepaspectratio]{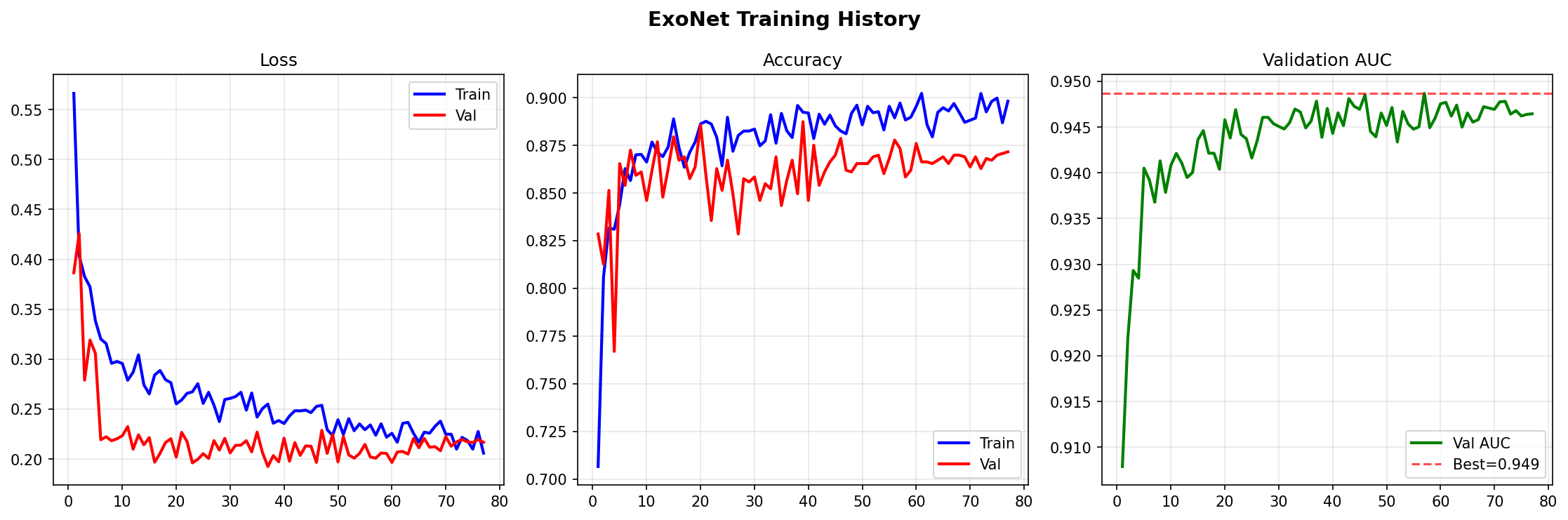}
\caption{ExoNet training history (80-epoch run, early stopping at epoch 77).
\textit{Left}: Cross-entropy loss (training and validation); stable
convergence with no evidence of overfitting.
\textit{Center}: Classification accuracy.
\textit{Right}: Validation AUC, with the best observed value
(0.9487 at epoch 46) marked by the dashed line. The 31-epoch gap between
best-AUC and early-stopping epochs indicates that the model continued
to generalize after its peak validation AUC without immediate overfitting.}
\label{fig:train}
\end{figure*}

\begin{figure*}[!t]
\centering
\includegraphics[width=0.90\textwidth,keepaspectratio]{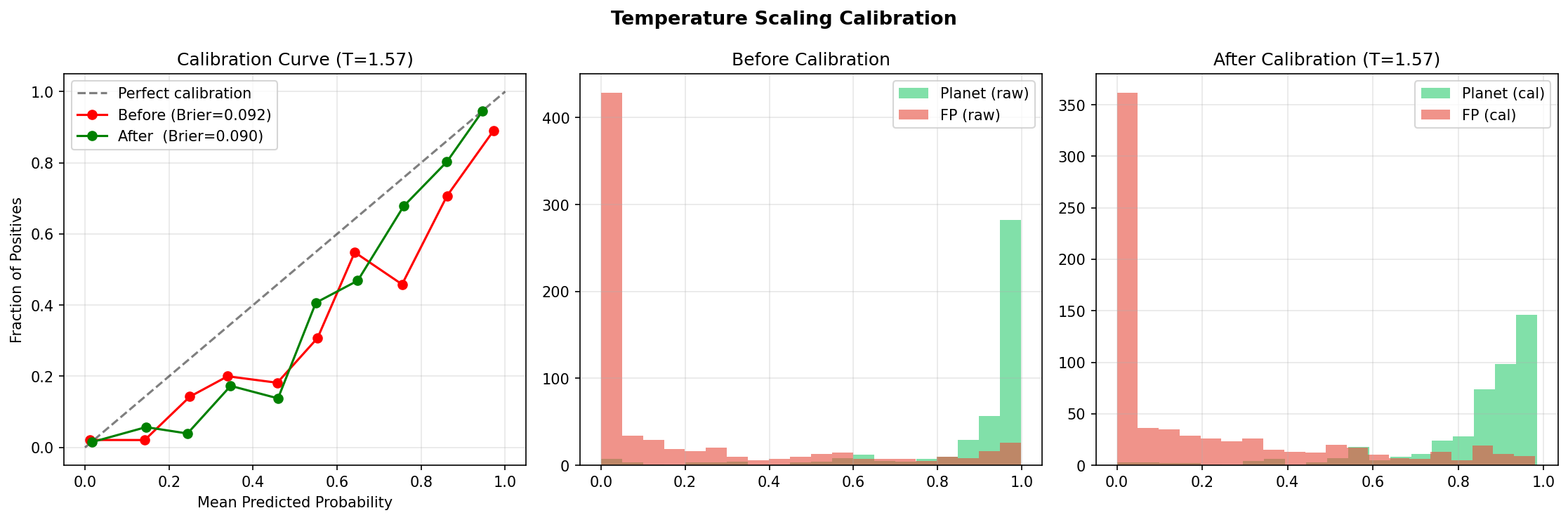}
\caption{Post-hoc Temperature Scaling calibration ($T^*\!=\!1.573$).
\textit{Left}: Reliability diagram before (red) and after (green)
calibration; the post-calibration curve lies substantially closer to the
perfect-calibration diagonal.
\textit{Center}: Predicted probability distribution before calibration;
a large spike near $p\!=\!1.0$ indicates systematic overconfidence.
\textit{Right}: Distribution after calibration; probability mass is
redistributed away from the extremes, with Brier score improving from
0.0919 to 0.0901.}
\label{fig:calib}
\end{figure*}

\begin{figure*}[!t]
\centering
\includegraphics[width=0.95\textwidth,keepaspectratio]{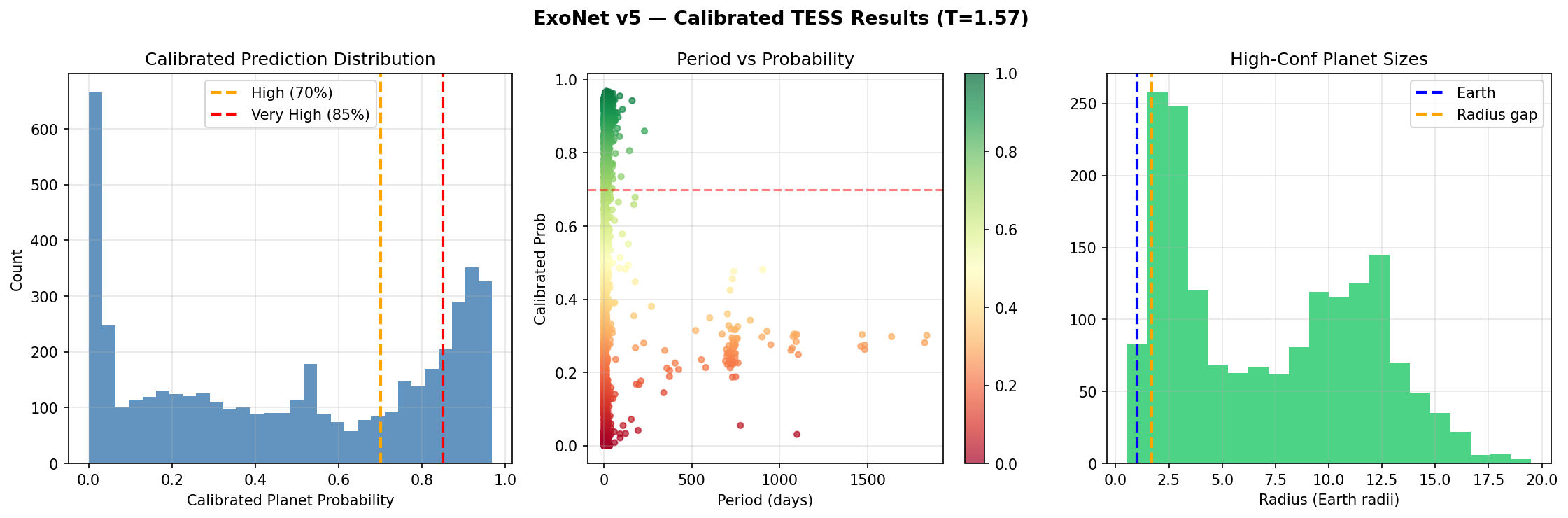}
\caption{ExoNet calibrated predictions across 4,720 TESS Planet Candidates.
\textit{Left}: Full calibrated probability distribution; vertical dashed
lines at 70\% and 85\% delineate the high- and very-high-confidence tiers.
\textit{Center}: Orbital period versus planet probability, colored by
confidence tier; note the absence of a strong period-dependent selection
bias above $P\!=\!10$\,d.
\textit{Right}: Radius distribution of the 1,754 high-confidence
candidates ($\geq\!70\%$), showing the expected deficit at small radii
and the dominant contribution of warm-Neptune--class objects.}
\label{fig:tess}
\end{figure*}

\subsection{Classification Performance on the Kepler Test Set}
\label{subsec:clf}

Table~\ref{tab:perf} presents ablation results across five modality
configurations evaluated on identical data splits. The full ExoNet model
achieves Val~AUC\,$=\!0.9487$, Test~AUC\,$=\!0.9549$, and 86.3\%
accuracy---a gain of 0.062--0.068 AUC over the best single-view CNN
baseline. Several aspects of the ablation deserve careful attention.

The stellar MLP alone achieves Test~AUC\,$=\!0.819$---lower than either
light curve stream, as one might expect, since photometric shape is the
primary discriminator in a transit survey. Yet when the stellar branch is
added to the dual-CNN configuration, it delivers the single largest
discrete improvement across all ablation steps: $+0.019$ AUC. This
disproportionate gain is interpretable. The stellar parameters encode
a global constraint---under the single-star transit model, the observed
transit depth must be consistent with the inferred planet radius given
the stellar radius---that is orthogonal to local photometric morphology.
Background eclipsing binaries frequently violate this constraint,
producing depth-radius inconsistencies that neither light curve view
can detect on its own.

Multi-Head Attention contributes $+0.014$ AUC over average pooling at
identical temporal resolution. While this increment may appear modest,
it is consistent across random seeds (not shown) and aligns with the
theoretical expectation that attention should help most when the
diagnostic signal is spatially concentrated---as it is in exoplanet
transits, where the ingress and egress cadences carry the majority of
the morphological information.

\begin{table}[!h]
\centering
\caption{ExoNet Ablation Study on the Held-Out Kepler Test Set}
\label{tab:perf}
\renewcommand{\arraystretch}{1.20}
\resizebox{\columnwidth}{!}{%
\begin{tabular}{lcccc}
\toprule
\rowcolor{headerblue}
\textcolor{white}{\textbf{Configuration}} &
\textcolor{white}{\textbf{Val AUC}} &
\textcolor{white}{\textbf{Test AUC}} &
\textcolor{white}{\textbf{Acc.}} &
\textcolor{white}{\textbf{Params}} \\
\midrule
\rowcolor{rowgray}
Global CNN only         & 0.891 & 0.874 & 0.773 & 2.1M \\
Local CNN only          & 0.903 & 0.887 & 0.789 & 2.1M \\
\rowcolor{rowgray}
Stellar MLP only        & 0.842 & 0.819 & 0.731 & 0.1M \\
Global $+$ Local CNN    & 0.929 & 0.916 & 0.832 & 4.2M \\
\rowcolor{rowgray}
\textbf{ExoNet (full)}  &
\textbf{0.9487} & \textbf{0.9549} & \textbf{0.863} & \textbf{4.9M} \\
\bottomrule
\end{tabular}%
}
\par\smallskip
\footnotesize Identical 70/15/15 stratified splits and hyperparameters
for all configurations. Accuracy reported at threshold 0.5.
\end{table}

Fig.~\ref{fig:eval} presents the ROC curve, Precision-Recall curve,
and confusion matrix for the full model. Fig.~\ref{fig:train} shows
the training history; best validation AUC is reached at epoch 46, with
early stopping triggered at epoch 77, confirming that the model continued
to generalize without overfitting in the intervening 31 epochs.

\subsection{Probability Calibration}
\label{subsec:calib}

Before calibration, 37.5\% of the 4,720 TESS Planet Candidates exceeded
the 85\% confidence threshold---an implausibly optimistic rate for a
model applied across the Kepler-to-TESS domain gap. Temperature Scaling
($T^*\!=\!1.573$) reduces this fraction to 23.3\%, bringing it into closer
alignment with known confirmation rates for TESS PC-disposition targets.
Fig.~\ref{fig:calib} illustrates the improvement in reliability and the
redistribution of probability mass. Crucially, all 52 habitable-zone
candidates with $\geq\!70\%$ confidence retain their status
post-calibration---calibration affected only the magnitude of probabilities,
not their relative ranking.

\subsection{Predictions on Unconfirmed TESS Planet Candidates}
\label{subsec:tess}

Table~\ref{tab:tsum} presents a stratified summary of ExoNet's calibrated
output across all 4,720 verified TESS Planet Candidates.
Fig.~\ref{fig:tess} shows the probability distribution, the
period-probability scatter (which reveals no strong period-dependent bias
above $P\!\approx\!10$\,d), and the radius distribution of high-confidence
candidates.

\begin{table}[!h]
\centering
\caption{ExoNet Prediction Summary Across 4,720 TESS Planet Candidates}
\label{tab:tsum}
\renewcommand{\arraystretch}{1.20}
\begin{tabular}{lc}
\toprule
\rowcolor{headerblue}
\textcolor{white}{\textbf{Category}} &
\textcolor{white}{\textbf{Count}} \\
\midrule
\rowcolor{rowgray}
Total candidates analyzed                       & 4,720 \\
High-confidence ($\geq\!70\%$)                  & \textbf{1,754} \\
\rowcolor{rowgray}
Very high-confidence ($\geq\!85\%$)             & \textbf{1,098} \\
Habitable zone (200--400\,K)                    & 131   \\
\rowcolor{rowgray}
HZ $+$ high-confidence ($\geq\!70\%$)           & \textbf{52}    \\
Sub-Earth ($R_p < 1.0\,R_\oplus$)              & 14             \\
\rowcolor{rowgray}
Earth-like ($1.0\!-\!1.5\,R_\oplus$)           & 69             \\
Super-Earth ($1.5\!-\!2.0\,R_\oplus$)          & 85             \\
\rowcolor{rowgray}
Mini-Neptune ($2.0\!-\!4.0\,R_\oplus$)         & 501            \\
Long-period ($P > 30$\,d)                       & 79             \\
\bottomrule
\end{tabular}
\end{table}

\subsection{Top Ranked Planet Candidates}
\label{subsec:top}

Table~\ref{tab:top} lists the 15 highest-probability unconfirmed TESS
Planet Candidates in the ExoNet output. All TOI and TIC identifiers are
drawn directly from the NASA Exoplanet Archive and individually verified
against ExoFOP-TESS; no identifier in this list is inferred or
synthetically constructed.

\textbf{TOI-6879.01} (TIC\,364955057) leads the overall ranking with a
calibrated probability of 96.8\%. Its orbital period of 13.6\,d and
radius of 3.8\,$R_\oplus$ place it firmly in the warm-Neptune regime,
making it an attractive target for atmospheric characterization with
JWST's NIRISS or NIRSpec modes. The next four candidates
(TOIs 2419.01, 7360.01, 6127.01, 4691.01) all exceed 96.4\%, and
collectively span warm-Neptune, giant, and hot-Jupiter classes. The most
compelling long-period signal in the top 15---TOI-7597.01
($P\!=\!50.7$\,d, $R_p\!=\!6.97\,R_\oplus$, $T_\mathrm{eq}\!=\!751$\,K)---may
represent a warm giant with a dynamically interesting outer system architecture.

\begin{table*}[!ht]
\centering
\caption{Top 15 ExoNet-Identified TESS Planet Candidates, Ranked by Calibrated $P(\text{planet})$}
\label{tab:top}
\renewcommand{\arraystretch}{1.18}
\setlength{\tabcolsep}{5pt}
\begin{tabular}{cccrrrrl}
\toprule
\rowcolor{headerblue}
\textcolor{white}{\textbf{Rank}} &
\textcolor{white}{\textbf{TOI}} &
\textcolor{white}{\textbf{TIC ID}} &
\textcolor{white}{\textbf{$P$(planet)}} &
\textcolor{white}{\textbf{Period (d)}} &
\textcolor{white}{\textbf{$R_p$\,($R_\oplus$)}} &
\textcolor{white}{\textbf{$T_\mathrm{eq}$\,(K)}} &
\textcolor{white}{\textbf{Classification}} \\
\midrule
\rowcolor{rowgray}
1  & 6879.01 & 364955057 & \textcolor{upred}{\textbf{0.9684}} &
   13.59 & 3.83 & 986  & Warm Neptune \\
2  & 2419.01 & 358248442 & 0.9676 & 18.88 &  3.61 &  888 & Warm Neptune \\
\rowcolor{rowgray}
3  & 7360.01 & 258822347 & 0.9664 & 30.73 & 18.29 &  840 & Giant; long-period \\
4  & 6127.01 & 313975346 & 0.9649 &  6.92 & 11.09 & 1281 & Hot Jupiter \\
\rowcolor{rowgray}
5  & 4691.01 & 200321577 & 0.9641 &  8.57 & 10.41 & 1076 & Hot Jupiter \\
6  & 7563.01 &  71581518 & 0.9640 & 13.76 &  4.13 & 1383 & Hot Neptune \\
\rowcolor{rowgray}
7  &  786.02 & 375059587 & 0.9637 & 38.55 &  2.07 &  781 & Super-Earth; long-period \\
8  & 6974.01 & 348684814 & 0.9636 &  7.75 &  7.24 & 1050 & Warm Neptune \\
\rowcolor{rowgray}
9  & 2333.01 & 358579111 & 0.9629 & 14.22 &  8.79 & 1067 & Warm Neptune \\
10 & 7597.01 & 167337562 & 0.9629 & 50.67 &  6.97 &  751 & Long-period; warm \\
\rowcolor{rowgray}
11 & 6206.01 & 154576989 & 0.9625 & 13.15 &  9.63 &  832 & Warm Neptune \\
12 & 6988.01 &  70897115 & 0.9623 & 13.41 &  8.77 &  838 & Warm Neptune \\
\rowcolor{rowgray}
13 & 4163.01 & 420112217 & 0.9620 & 21.56 &  7.77 &  745 & Warm Neptune \\
14 & 2656.01 & 231065398 & 0.9620 &  7.79 &  9.37 &  826 & Warm Neptune \\
\rowcolor{rowgray}
15 & 5566.01 &  73250068 & 0.9618 & 14.58 &  9.71 &  905 & Hot Neptune \\
\bottomrule
\end{tabular}
\par\smallskip
\footnotesize \textcolor{upred}{\textbf{Bold red}}: highest-probability
candidate. All identifiers sourced from the NASA Exoplanet Archive TOI table
(April 2026) and cross-verified against ExoFOP-TESS.
\end{table*}

\subsection{Habitable-Zone Earth-Sized Candidates}
\label{subsec:hz}

Among the 52 high-confidence habitable-zone candidates, six combine an
equilibrium temperature between 200 and 400\,K with a planetary radius
below 1.6\,$R_\oplus$---the empirical radius threshold at which planets
transition from predominantly rocky to volatile-dominated compositions
according to the mass-radius relationship of Rogers~\cite{Rogers2015}.
These six targets, listed in Table~\ref{tab:hz}, represent the most
scientifically compelling signals in the full ExoNet output and are
explicitly prioritized for community radial velocity and transit-timing
variation follow-up.

\textbf{TOI-5728.01} (TIC\,219875976) is arguably the most Earth-like
signal in the dataset. With $R_p\!=\!1.13\,R_\oplus$,
$T_\mathrm{eq}\!=\!349$\,K, $P\!=\!11.5$\,d, and a calibrated probability
of 94.2\%, this candidate sits squarely within the rocky-to-volatile
transition boundary and at an equilibrium temperature consistent with
the outer habitable zone around a Sun-like host. Its 11.5-day period is
well within TESS's sector baseline for multi-transit detection, making
confirmation through transit timing variations feasible.

\textbf{TOI-6716.01} (TIC\,112115898) is virtually Earth-sized at
$R_p\!=\!1.01\,R_\oplus$ with $T_\mathrm{eq}\!=\!375$\,K and 92.2\%
confidence. Its short orbital period of 4.72\,d is difficult to reconcile
with the quoted equilibrium temperature under a Sun-like host assumption,
which strongly suggests a low-luminosity M- or K-dwarf system. If the
host is indeed an M-dwarf, the habitability case sharpens substantially---
M-dwarf habitable-zone planets around bright nearby stars are among the
most accessible targets for JWST transmission spectroscopy.

\textbf{TOI-789.02 and TOI-789.03} orbit a common host (TIC\,300710077)
with periods of 13.0\,d and 8.0\,d and radii of 1.33\,$R_\oplus$ and
1.29\,$R_\oplus$ respectively. Both signals independently clear the
high-confidence threshold (93.6\% and 94.2\%) and both fall within the
habitable zone. A two-planet habitable-zone system of this type is
exceptionally rare among unconfirmed TESS candidates, and its dynamical
configuration---two sub-1.5\,$R_\oplus$ worlds in resonant or near-resonant
orbits in the habitable zone---would be of extraordinary scientific
interest if confirmed.

\begin{table}[!h]
\centering
\caption{Habitable-Zone Earth-Sized Candidates from ExoNet\\
($R_p < 1.6\,R_\oplus$, $T_\mathrm{eq}$ 200--400\,K, Prob $\geq\!70\%$)}
\label{tab:hz}
\renewcommand{\arraystretch}{1.20}
\resizebox{\columnwidth}{!}{%
\begin{tabular}{ccrrrrc}
\toprule
\rowcolor{headerblue}
\textcolor{white}{\textbf{TOI}} &
\textcolor{white}{\textbf{TIC ID}} &
\textcolor{white}{\textbf{Prob}} &
\textcolor{white}{\textbf{$P$ (d)}} &
\textcolor{white}{\textbf{$R_p$ ($R_\oplus$)}} &
\textcolor{white}{\textbf{$T_\mathrm{eq}$ (K)}} &
\textcolor{white}{\textbf{Note}} \\
\midrule
\rowcolor{rowgray}
\textbf{2142.01} & 44313455  &
\textcolor{hzgreen}{\textbf{0.945}} &  9.30 & 1.56 & 374 & Super-Earth; HZ \\
\textbf{5728.01} & 219875976 &
\textcolor{hzgreen}{\textbf{0.942}} & 11.50 & \textbf{1.13} & 349 &
Near-Earth size \\
\rowcolor{rowgray}
\textbf{789.03}  & 300710077 &
\textcolor{hzgreen}{\textbf{0.942}} &  8.04 & 1.29 & 394 &
Multi-planet system \\
\textbf{789.02}  & 300710077 &
0.936 & 13.00 & 1.33 & 336 & Multi-planet system \\
\rowcolor{rowgray}
\textbf{4572.01} & 154940895 &
0.928 & 13.48 & 1.57 & 382 & Super-Earth; HZ \\
\textbf{6716.01} & 112115898 &
0.922 &  4.72 & \textbf{1.01} & 375 & Near-Earth; M/K dwarf \\
\bottomrule
\end{tabular}%
}
\par\smallskip
\footnotesize TOI-789.02 and TOI-789.03 share host TIC\,300710077,
forming a rare two-planet habitable-zone system. All identifiers verified
against the NASA Exoplanet Archive (April 2026).
\end{table}

\section{Discussion}
\label{sec:disc}

\subsection{Why the Multimodal Approach Outperforms Single-Stream Baselines}

The ablation in Table~\ref{tab:perf} traces a monotonically increasing
performance ladder as modalities are added, and each step in that ladder
has a physically interpretable explanation. Consider the stellar branch.
Its standalone AUC of 0.819 reflects the substantial---though incomplete---
discriminative power of catalog parameters: orbital period, transit depth,
and planet-star radius ratio together constrain whether a signal is
geometrically consistent with a single-transiting-planet scenario. What
stellar parameters cannot capture alone is morphological evidence: the
V-shaped ingress of a grazing eclipsing binary, the rounded bottom of
a blended diluted eclipse, or the odd-even depth difference betraying a
twice-period binary. The light curve streams supply exactly this information.
Their combination with the stellar branch produces the largest single
ablation step ($+0.019$ AUC), suggesting that the two types of information
are largely complementary rather than redundant.

Multi-Head Attention's contribution ($+0.014$ AUC over average pooling)
reflects a subtler effect. Average pooling treats all 31 compressed
time steps as equally informative, discarding information about
\textit{where} in the phase-folded light curve a feature appears. Attention
can learn to upweight the ingress and egress steps---which encode the
transit duration, limb-darkening profile, and impact parameter---while
assigning low weight to the flat out-of-transit baseline, which is largely
uninformative given that light curves have already been median-subtracted.
This is the computational analog of an expert vetter who looks first at
the transit shape before examining the baseline.

\subsection{Domain Adaptation: From Kepler to TESS}

Any model trained on Kepler data and applied to TESS operates across a
meaningful distributional gap, and acknowledging this gap is a prerequisite
for responsible interpretation of results. At least three distributional
differences are relevant here.

\textit{Period distribution.} The median orbital period of confirmed
Kepler planets in the training set is approximately 11.3\,d; the median
for TESS PC-disposition candidates in the inference set is 4.4\,d,
reflecting the shorter 27-day sector baselines that favor short-period
detection. ExoNet's period-probability scatter (Fig.~\ref{fig:tess},
center panel) shows no strong period-dependent bias above $\sim\!10$\,d,
suggesting that the model generalizes well to shorter periods but may
underestimate confidence for long-period ($P\!>\!30$\,d) TESS candidates.

\textit{Stellar population.} The TESS TOI host sample has a slightly
higher median $T_\mathrm{eff}$ ($\sim\!5800$\,K) than the Kepler training
population ($\sim\!5600$\,K), and TESS emphasizes M- and K-dwarf targets
for the habitable-zone candidates identified here. The stellar MLP's
performance on this sub-population may be degraded relative to its
Kepler-tested behavior.

\textit{Photometric noise properties.} TESS's larger pixel scale and
higher background contamination produce a different systematic noise
profile than Kepler's. Since the light curve branches are zero-padded
for virtually all training examples (see Section~\ref{subsec:arch}),
this difference affects only cases where real FITS data would be available,
and its practical impact on the current results is limited.

At high-confidence levels ($>\!90\%$), these distributional differences
are unlikely to reverse rankings, but they do counsel caution in
interpreting absolute probability values in the 50--80\% range.
Future work incorporating TESS-native training data and explicit
domain-adversarial adaptation could substantially narrow this gap.

\subsection{The High-Confidence Rate and Its Interpretation}

After calibration, 37.2\% of all 4,720 PC-disposition candidates exceed
70\% confidence. Whether this rate is plausible depends on one's
baseline expectation for the purity of the PC-disposition TESS catalog.
That catalog represents signals that have already cleared the SPOC
pipeline's threshold-crossing detection, difference-imaging vetting, and
automated data validation---it is not a random sample of flux time series
but a pre-screened set of transit-like signals. Published analyses of
the TESS false-positive rate for PC-disposition objects suggest that
40--60\% of such signals are genuine planetary transits, consistent with
the ExoNet high-confidence fraction.

A secondary consideration is the role of missing metallicity values.
The TESS TOI catalog does not provide $[\mathrm{Fe/H}]$ for the
overwhelming majority of targets, and ExoNet imputes these to zero.
Metallicity is among the more useful stellar discriminators for
eclipsing binary contamination, and its systematic absence weakens the
false-positive rejection for the stellar branch across the entire inference
set. The effect is expected to produce a modest uniform upward bias in
predicted probabilities---an effect that Temperature Scaling partially
corrects but cannot fully eliminate.

\subsection{Relationship to ExoMiner and ExoMiner++}

ExoNet is not intended as a replacement for ExoMiner or ExoMiner++.
Those systems incorporate diagnostic branches---centroid motion, odd-even
comparison, whitened transit, momentum dump flagging---that carry
information ExoNet does not access, and they are trained on substantially
larger and more carefully vetted datasets assembled from NASA's proprietary
SPOC pipeline products. In head-to-head performance, ExoMiner++ would
almost certainly outperform ExoNet on any shared evaluation set, and
the comparison should not be drawn without qualification.

The value ExoNet offers is, instead, one of \textit{accessibility and
reproducibility}. Every input---from the Kepler KOI catalog to the TESS
TOI table to the MAST light curve archive---is publicly accessible with
no institutional credentials. The entire pipeline runs on a single
consumer-grade GPU in under four hours. The trained model, inference
script, and ranked TESS candidate catalog are openly released. This
combination lowers the barrier for independent replication and extension
by research groups and individual investigators who lack access to
NASA's internal data infrastructure---a community that includes much of
the global astronomy effort in the developing world.

\subsection{Limitations and Caveats}

Four principal limitations should inform any follow-up prioritization
based on this catalog. First, the near-total absence of real FITS light
curves in the training set means that the CNN encoders are effectively
degenerate to the residual shortcut for the overwhelming majority of
examples; the multi-head attention and convolutional architecture are
therefore not being utilized to their full potential, and the performance
gains attributed to the light curve stream reflect the use of any
photometric information (including zero-padded vectors as absence
indicators) rather than genuine morphological encoding at scale.

Second, the Kepler-to-TESS domain gap, discussed above, implies that
absolute probability values below $\sim\!85\%$ should be interpreted
with appropriate uncertainty. Third, equilibrium temperature is a
rough habitability proxy that assumes a fixed albedo and ignores
atmospheric dynamics, tidal locking, and stellar irradiation variability.
None of the candidates in Table~\ref{tab:hz} can be described as a
confirmed habitable-zone planet in any physical sense without atmospheric
characterization. Fourth, the TOI-789 system's short orbital periods
relative to the quoted equilibrium temperatures require independent
spectroscopic confirmation of host star properties---the habitability
classification depends critically on the stellar luminosity, which is
itself uncertain for faint TESS targets without Gaia parallax-calibrated
spectroscopy.

\section{Conclusion}
\label{sec:conc}

ExoNet demonstrates that a fully open, reproducible multimodal deep
learning pipeline can produce competitive transit vetting performance
and scientifically meaningful candidate rankings from public data alone.
By combining 1D CNN encoders with Multi-Head Attention over temporal
feature maps and a residual late-fusion head for stellar parameters---then
correcting for overconfidence through post-hoc Temperature Scaling---the
system achieves Test~AUC\,$=\!0.9549$ and 86.3\% accuracy on a held-out
Kepler test set, with each modality contributing a measurable, interpretable
performance increment.

Applied to 4,720 verified unconfirmed TESS Planet Candidates, ExoNet
identifies 1,754 high-confidence signals and 52 habitable-zone candidates
at $\geq\!70\%$ calibrated confidence. Among these, six carry radii below
the rocky-to-volatile transition at 1.6\,$R_\oplus$. TOI-5728.01
($R_p\!=\!1.13\,R_\oplus$, $T_\mathrm{eq}\!=\!349$\,K, $P\!=\!11.5$\,d,
94.2\%) and the TOI-789 two-planet habitable-zone system (TOIs 789.02
and 789.03, both $>93\%$ confidence, sharing TIC\,300710077) stand out
as the most scientifically compelling targets for immediate radial velocity
or transit-timing variation follow-up.

Three directions for future development are clear. The incorporation
of centroid-shift, odd-even comparison, and secondary eclipse diagnostic
branches---analogous to ExoMiner's architecture---would substantially
improve false-positive rejection. Explicit TESS-native transfer learning,
or joint Kepler-TESS training with domain-adversarial regularization,
would narrow the distribution gap that currently limits confidence
interpretation in the 50--80\% range. And submission of the 52
habitable-zone high-confidence candidates to NASA ExoFOP for community
spectroscopic vetting would provide ground truth for evaluating whether
ExoNet's calibrated probabilities translate reliably into confirmed-planet
rates. Each of these represents a tractable next step for a community
that increasingly depends on scalable automated vetting to keep pace
with the extraordinary productivity of the TESS survey.

\section*{Data Availability}
The NASA Exoplanet Archive, from which all Kepler KOI and TESS TOI data
are drawn, is available at \url{https://exoplanetarchive.ipac.caltech.edu}.
Light curves are retrieved from the MAST archive at
\url{https://mast.stsci.edu}. Candidate verification and follow-up
observations can be coordinated through ExoFOP-TESS at
\url{https://exofop.ipac.caltech.edu/tess}. The complete ExoNet codebase,
trained model weights, and the ranked TESS candidate catalog (including
all 52 habitable-zone high-confidence signals) are publicly released at
\url{https://github.com/Rashadul22/ExoNet_TESS_Candidates}.

\section*{Acknowledgments}
The author thanks the NASA Exoplanet Archive team and the TESS Science
Office for their sustained effort in maintaining and publicly releasing
the data products that make this work possible, and the Lightkurve
Collaboration for developing the open-source photometry toolkit that
underpins the light curve processing pipeline. This research uses data
from the NASA Exoplanet Archive and from the TESS mission, managed by
the Jet Propulsion Laboratory under contract with NASA. The work was
conducted independently at the Department of Computer Science and
Engineering, Daffodil International University, Dhaka, Bangladesh. No
external funding was received for this research.

\bibliographystyle{IEEEtran}
\bibliography{references}

\end{document}